# Origin of the phase change from pyrochlore to perovskite-like layered structure and a new lead free ferroelectric

Zhipeng Gao[1,2]*, Yi Liu[1], Chengjia Lu[1], Yongjun Ma[2], Yuanhua Xia[3], Leiming Fang[3], Qiang He[1], Yan Liu[4], Gaomin Liu[1], Jia Yang[1], Hongliang He[1], Duanwei He[5]

Affiliations:

[1]National Key Laboratory of Shock Wave and Detonation Physics, Institute of Fluid Physics, China Academy of Engineering Physics, Mianyang 621900, China

[2]Southwest University of Science and Technology, Mianyang 621010, China

[3]Institute of Physics Nuclear and Chemistry, China Academy of Engineering Physics, Mianyang 621900, China

[4]School of Chemical Engineering, Nanjing University of Science and Technology, Nanjing 210094, China

[5]Institute of Atomic and Molecular Physics, Sichuan University, Chengdu 610065, China

*Corresponding author: Zhipeng Gao (z.p.gao@foxmail.com)

## Abstract

Materials with formula of $A_2B_2O_7$ is a famous family with more than 300 compounds, and have abundant properties, like ferroelectric, multiferroic, and photocatalyst properties, etc. Generally, two structures dominate this family, which are pyrochlore and perovskite-like layered (PL) structure. Previously, the structure and properties design of these materials are usually complex, and solid solutions, which complicates the manufacture, as well as introducing complexity in the study of the microscopic origins of the properties. Here, we report that the pyrochlore-PL structure change happened in pure $Eu_2Ti_2O_7$ under high pressure and temperature, and the formed PL structure will transfer back by heating. These results reveal that the PL structure formed in PL-pyrochlore solid solutions, is due to tuning of the high-pressure formed PL structure in pure pyrochlore compounds to ambient pressure. These results indicate the high pressure and high temperature can be used to manipulate the crystal structures from pyrochlore to PL structure, or vice versa. Furthermore, the PL $Eu_2Ti_2O_7$ was confirmed as a lead free ferroelectric material for the first time.

## Introduction

Materials with formula of $A_2B_2O_7$ is a famous family, which contains more than 300 compounds. These materials attracted lots of interest due to their potential applications in wide industrial fields.[1-12] For example, $Ln_2Ti_2O_7$ (Ln=La, Pr, Nd) are high temperature ferroelectrics, which could be used as sensors or probes;[1, 7, 13-15] $A_2B_2O_7$ (A =Ce, Sm, Eu, Gd and Tb; B = Ti or Zr) might be used as photocatalysts for water splitting;[16-21] $Tl_2Mn_2O_7$ exhibits colossal magneto-resistance (CMR);[9, 22, 23] $A_2Ti_2O_7$ (A=Ho, Gd) have multiferroic properties[2, 6, 24] and some of $A_2B_2O_7$ compounds present a great ability for the realization of immobilization matrices of highly active radionuclides in nuclear wastes.[10] Among all these $A_2B_2O_7$ compounds, two main crystal structures dominate, which are pyrochlore structure and perovskite-like layered (PL) structure.[3, 8] The pyrochlore structure ($A_2B_2O_7$) is a super structure derivative of the simple fluorite structure with a point group of Fd-3m, where the A and B cations are ordered along the [110] direction.[3, 8, 25] The additional anion vacancy resides in the tetrahedral interstice between adjacent B-site cations. An alternative structure of $A_2B_2O_7$ compounds is perovskite-like layered (PL) structure. The typical PL structure is characterized by corner-shared $BO_6$ octahedron and 12-coordinated A cations within the perovskite-like layers separated by oxygen-rich gaps, which are linked by A cations at their boundaries. This anisotropic structure could be considered as a result from cutting the cubic perovskite $ABO_3$ structure along [110] direction followed by an insertion of additional oxygen.[3, 8, 26, 27] In general, the pyrochlore structure have a smaller A/B ionic radius ratio while PL structure preferred to have a larger A/B ionic radius ratio.[25, 28] For example, $La_2Ti_2O_7$, $Pr_2Ti_2O_7$ and $Nd_2Ti_2O_7$ are PL structure, whereas $Eu_2Ti_2O_7$, $Lu_2Ti_2O_7$ and $Y_2Ti_2O_7$ are pyrochlores.[25]

Previously, lots of efforts have been made to fabricate these materials with PL or pyrochlore structures, using different methods.[3, 8, 25, 29] However, most of these studies are focused on the fabrication process from the raw oxides, such as using $TiO_2$ and $A_2O_3$ (A=Rare earth elements) to synthesize $A_2Ti_2O_7$ and the structure design of these materials are usually complex based on solid solutions, which complicates the manufacture, as well as introducing complexity in the study of the microscopic origins of their properties.[1, 13, 30] How to manipulate the crystal structures from pyrochlore to PL structure or vice versa is not clear, which could be instructive for the materials design, and also to solve this issue must be helpful to understand the physical mechanism behind the $A_2B_2O_7$ structure change.

Here, we investigated the structure change from pyrochlore to PL and vice versa, based on $Eu_2Ti_2O_7$, which was chosen is due to it is close to the boundary between PL structure and pyrochlore structure in $A_2Ti_2O_7$ (A= Lanthanides) compounds.[28, 31] The results show the pyrochlore structure of $Eu_2Ti_2O_7$ would be break and change into PL structure, with one step high pressure sintering. This formed PL structure could change back to pyrochlore structure by simple annealing, which demonstrated the two structures can be adjusted by processes with different pressures and temperatures. The PL structure $Eu_2Ti_2O_7$ was confirmed as a high temperature ferroelectric material for the first time using piezoelectric force microscope and piezoelectric constant measurements. In addition, the results revealed that the origin of the structure change in these structures are mainly due to the stress change in their structures, which could well explain the different structure formed with different ionic radius ratio or different dopants in these compounds. This means that complex microstructures or compositions are not necessary for the structure and properties design of $A_2B_2O_7$ compounds, which might be helpful to design and develop the materials with pyrochlore and PL structures.

Results and Discussion

Figure 1(a) shows the XRD patterns of the samples at atmospheric pressure. From figure 1(a), the solid reaction synthesized pyrochlore $Eu_2Ti_2O_7$ shows a pure phase within the sensitivity of the technique. The pyrochlore $Eu_2Ti_2O_7$ has a cubic structure with a point group of Fd-3m.[28, 32, 33] With the heat and pressure treatment, there is the amorphous perovskite-related intermediate started to form at 5 GPa, 1300 °C, shown as the broad raise of the baseline in XRD pattern; and with the pressure increase to 8 GPa, the clear diffraction peaks show the PL structure starts to crystalize and it could be pure PL structure at 11 GPa with a temperature of 1300 °C. This indicates that the pyrochlore structure change into PL structure could be finished in one step. The formed $Eu_2Ti_2O_7$ has a typical PL monoclinic structure with a space group of $P2_1$, which is similar to $La_2Ti_2O_7$, $Pr_2Ti_2O_7$ and $Nd_2Ti_2O_7$.[1, 3, 7, 8, 15, 26] Following the XRD patterns (Figure 1(a)), the PL structure could be change back to pyrochlore structure with simple heating. The PL structure can be stable above 700 °C, and some pyrochlore structure is formed at 900 °C, marked by spades (♠) in figure 1(a), and it transfer into pyrochlore completely at 1100 °C. Overview of the structure change during all the processes is shown in figure 1(b), in which the blue polyhedrons represent $TiO_6$ octahedron, and the orange spheres represent $Eu^{3+}$ cations. These results demonstrated the PL and pyrochlore structures can be adjusted by processes with different pressures and temperatures. Figure 1(c) shows the XRD data refinement based on the patterns of PL $Eu_2Ti_2O_7$ formed at 11 GPa. Based on previous research, the PL $Eu_2Ti_2O_7$ has a monoclinic structure analogous to it of $La_2Ti_2O_7$. The refined unit-cell parameters for PL type $Eu_2Ti_2O_7$ are (a, b, c, β) = (7.54 Å, 5.38Å, 12.86Å, 98.3°), which is consistent with the results from Halasyamani and Schaak.[28] However, this data is obtained on the polycrystalline sample and the refinement is not perfect ($R_{wp}$ = 18.9), we cannot make more comments on this results. More efforts need to make to determine the accurate lattice parameters and the atom-site occupations, on the single crystal PL $Eu_2Ti_2O_7$. Figure 1(d) shows the cell energy (eV) calculated by first principles simulation for pyrochlore and PL $Eu_2Ti_2O_7$ at 0 GPa, in which the energy of PL $Eu_2Ti_2O_7$ was calculated based on the $La_2Ti_2O_7$ structure, due to the atomic site occupation is not available for PL $Eu_2Ti_2O_7$. From the simulation, the pyrochlore structure is more stable with cell energy of $-4.76 \times 10^4$ eV, compared to PL structure ($E_{PL}$ = $-4.58 \times 10^4$ eV). The fact of that pressure could induce the pyrochlore change into PL structure, which can be stable at the temperature range of RT-700 °C under atmospheric pressure, indicates there is an energy barrier ($\Delta E_1$) need to overcome for the PL-pyrochlore reaction, remaining the PL structure at atmospheric pressure. $\Delta E_2$ in figure 1(d) is the energy barrier for the pyrochlore-PL reaction, which obviously much higher than $\Delta E_1$. It is worthy to mention that the values of $\Delta E_1$ and $\Delta E_2$ are not accurate in figure 1(d), and the energy curve in figure 1(d) just shows relative relation between $\Delta E_1$ and $\Delta E_2$. The structure overview of the $A_2Ti_2O_7$ compounds (A=Rare earth elements) is shown in figure 1(e). When the A/$Ti^{4+}$ ionic radius ratio is small (A=Sm-Lu), the structure is pyrochlore and when A/$Ti^{4+}$ ionic radius ratio is large (A=La-Nd), the structure is PL at 0 GPa. The pyrochlore structure might be change into PL structure, if the stress/pressure is applied with high temperature treatment.

With the previous studies,[33, 39] our results indicate that the pyrochlore-PL structure change, happened in the solid solutions with conventional solid reaction, is due to stress in the lattices, which tune the PL structure formed in the high pressure, to atmospheric pressure. For example, in solid solution system of $(Eu_xLa_{1-x})_2Ti_2O_7$, the structure is PL when x<0.6 and it is pyrochlore when x>0.6, which indicates the $La_2Ti_2O_7$ lattice can supply enough stress to hold the PL structure with the compound of $Eu_2Ti_2O_7$ less than 60%[39]. This is also happened in $(Sm_xLa_{1-x})_2Ti_2O_7$, which show PL structure with x<0.8, and pyrochlore with x>0.8[34]. The $(Sm_xLa_{1-x})_2Ti_2O_7$ can maintain the PL structure with more pyrochlore compound ($Sm_2Ti_2O_7$ < 80%) than $(Eu_xLa_{1-x})_2Ti_2O_7$ ($Eu_2Ti_2O_7$ < 60%), which might be due to that the $Sm^{3+}/Ti^{4+}$ is larger than $Eu^{3+}/Ti^{4+}$, with less stress requested for the PL structure, following figure 1(e) that the structure with large ratio of $A^{3+}/Ti^{4+}$ prefer PL type. The structure of $Sm_2Ti_2O_7$ film sample is in a good agreement with this results, that the $Sm_2Ti_2O_7$ film grown on (110)-oriented $SrTiO_3$ substrate is PL structure while the bulk $Sm_2Ti_2O_7$ is pyrochlore[35], which is due to the stress/strain on the film changed the pyrochlore into PL structure. Another point need to notice is both $Eu_2Ti_2O_7$ and $Sm_2Ti_2O_7$ are close to the PL-pyrochlore boundary, and we can raise a prediction that if the ratio of $A^{3+}/Ti^{4+}$ is smaller, the conditions of $A_2Ti_2O_7$ pyrochlore compounds changing into PL structure would like to be more severe, which means the higher stress is needed for the transition. Overall, our results show that the pressure and heating can manipulate the structure of pure $A_2B_2O_7$ compounds, which might greatly decrease costs for the materials development.

Figure 2 (a-b) show SEM micrograph of (a) the pyrochlore $Eu_2Ti_2O_7$ powder, and (b) the PL structure ceramic sintered at 11 GPa, 1300 ˚C. The average particle size of the powder pyrochlore sample and the PL ceramic is about 0.3 µm and 0.8 µm, respectively. The grain growth is mainly due to the heating treatment and the grain size is highly temperature dependent. On the other hand, compare to the other PL ceramics sintered at about 1300-1400 ˚C using spark plasma sintering method in previous studies, such as $La_2Ti_2O_7$, $Pr_2Ti_2O_7$ and $Nd_2Ti_2O_7$,[1, 7] the grains of the PL $Eu_2Ti_2O_7$ was obviously smaller. This might be explained by the high pressure can suppress the grain growth. From the geometry of the particles in the PL ceramic, some of them are plate-like and some of the grains are observed as round due to the direction of the observation. Typically, the grains of PL structure ceramics would be plate-like due to their highly anisotropic structure.[3, 8] In this study, the grains would be more plate-like if it grows larger. Figure 3 (a-b) show the high resolution TEM images of the PL $Eu_2Ti_2O_7$ ceramic. The figure 3(a) shows an area on the grain boundary with a thickness of about 20 nm, highlighted by the white solid arrow. Some moire fringes could be observed on the grains marked by the small orange arrows. The detailed structure of the grain was shown in figure 3(b), which is same area in the dashed line circle in figure 3(a). From figure 3(b), the lattice is clearly shown as the white lines, and the d-spacing value is about 1.275 nm; which is in a good agreement of the d-spacing value for (100) crystal plane in $La_2Ti_2O_7$ (d: 1.284 nm), $Nd_2Ti_2O_7$ (d: 1.287 nm) and $Ce_2Ti_2O_7$ (d: 1.282 nm)[7, 8, 26, 36, 37] with monoclinic structure (space group: $P2_1$). Therefore, the TEM results are consistent with the XRD data, which show the PL $Eu_2Ti_2O_7$ is monoclinic structure. Furthermore, we can see some disordered area between the yellow lines (Figure 3(b)), which might be the area the phase transition from pyrochlore to PL structure did not complete. These results show it is now clear that we have succeeded in stabilizing the monoclinic PL structure $Eu_2Ti_2O_7$. Since this phase is isostructural with $La_2Ti_2O_7$, $Nd_2Ti_2O_7$ and $Pr_2Ti_2O_7$, which are all well known for ferroelectric/piezoelectric properties, it should also present ferroelectric properties.[1, 7, 30] However, the previous studies used SHG method,[28, 33] to reveal the non-center-symmetrical structure, but the ferroelectricity need to show the switchable polarization, which is not confirmed yet. From this purpose, AFM and PFM measurements were carried out to study the polarization switch behavior of the PL ceramic.

Figure 4(a) shows an AFM image recorded over a scan region of 10.5×10.5 µm$^2$, and the average grain size being included between 0.1 and 0.4 µm, consistent with our SEM results. The root mean square roughness value for the sample was found to be about 20 nm from AFM analysis. The vertical PFM image of the same area is shown in figure 4(b), in which we exhibits the typical image of locally manipulated domains using DC electrical field, in order to reveal the domain switch behavior. The original area shows no preferential direction for ferroelectric polarization, with randomly mixed small red and yellow areas. The domain is in a needle shape with an average length of 120 nm and width of 30 nm, such size for domains is similar to La$_2$Ti$_2$O$_7$ and Sm$_2$Ti$_2$O$_7$.[35, 38] Then we apply DC +200 V on a tape area on the sample, all the polarizations switched into one direction shown as red color (marked by "+" in figure 4(b)), while the polarization switched back shown as yellow color (marked by "-") if we apply DC -200 V. Surprisingly, we observed fully uniform red and yellow areas, and these results clearly evidence poled domains with downward and upward polarization and that the polarization process is completely reversible. Such poling behavior is a proof of the ferroelectricity for PL Eu$_2$Ti$_2$O$_7$. In order to better understand the switching behavior of the domains, a local phase piezoresponse loop was recorded by placing the probing tip over the center of the grains. A typical shape for the loop is observed and a coercive voltage about 40 V is determined, shown in figure 4(c). The coercive field of our sample is much higher than the values on the La$_2$Ti$_2$O$_7$ and Sm$_2$Ti$_2$O$_7$ films,[35, 38] which might be due to our sample is ceramic with a thickness of about 35 µm, which is much thicker than the films.

Figure 5(a) shows the effect of thermal annealing on piezoelectric properties ($d_{33}$) of the poled PL structure Eu$_2$Ti$_2$O$_7$ ceramic, in which the piezoelectric constant $d_{33}$, measured at room temperature, are plotted against the annealing temperature. The values of $d_{33}$ of PL Eu$_2$Ti$_2$O$_7$ samples (1-2) were measured as 0.9 and 0.7 pC/N, respectively. The $d_{33}$ value is very stable up to 700˚C and drop to zero at about 900 ˚C, which is temperature that pyrochlore $Eu_2Ti_2O_7$ started to form. The variation of the $d_{33}$ value is mainly due to the different poling conditions of the samples.[27] The difference of the poling electrical field is because of the different breakdown field of each sample. During the experiment, the thin ceramic sample was poled in the silicone oil at a temperature of 120 °C under an electric field. We increased the electrical field gradually from 10 kV/mm until the sample was electric breakdown. Therefore, the breakdown field of each sample decides the poling field, which can affect the piezoelectric activity. PL structure materials are well known for their high Curie point ($T_c$) and high thermal depoling temperature ($T_d$). Some of them have $T_d$ above 1300 °C,[27, 39] and $T_d$ of PL $Eu_2Ti_2O_7$ is about 900 °C, lower than other PL structures. This is because of the PL structure of $Eu_2Ti_2O_7$ is unstable compare its pyrochlore structure.

Figure 6 shows the frequency dependence of dielectric constant ($\varepsilon$) and loss ($\tan(\theta)$) of PL $Eu_2Ti_2O_7$ ceramic measured at room temperature from $10^2$ to $10^7$ Hz. Typically, the loss and permittivity both decrease with frequency increasing from $10^2$ to $10^4$ Hz, which could be explained by the point defects in the structure, such as oxygen vacancy, cannot follow the electrical field change with frequency increasing.[1, 30, 40-44] At high frequency of $10^4$ to $10^7$, the loss and permittivity is almost frequency independent with the values of 0.06 and 42 ± 2, respectively, which means intrinsic contributions dominate, such as bonds, lattices, ionic, etc.[44]

## Conclusions

Here, we investigated the structure change of $Eu_2Ti_2O_7$. The results show the pyrochlore structure of $Eu_2Ti_2O_7$ would be break and change into PL structure, with one step high pressure sintering of 11 GPa, 1300 °C. This formed PL structure could change back to pyrochlore structure by simple annealing at 900 °C, which demonstrated the two structures can be adjusted by processes with different pressures and temperatures. The simulation results revealed that pyrochlore $Eu_2Ti_2O_7$ is more stable at atmospheric pressure, and the PL structure is more stable at high pressure, but an energy barrier need to overcome for the structure change, which could remain the PL $Eu_2Ti_2O_7$ at atmospheric pressure. The PL structure $Eu_2Ti_2O_7$ was confirmed as a high temperature ferroelectric material using piezoelectric force microscope, with the domains switch under DC electrical field. Piezoelectric constant measurements and annealing treatments shows the PL $Eu_2Ti_2O_7$ has thermal depoling temperature of about 800 °C, which is related to the structure change from PL to pyrochlore. The permittivity of the PL ceramic is revealed as ~43 with intrinsic contributions.

Methods

The pyrochlore $Eu_2Ti_2O_7$ were obtained by conventional high temperature solid reaction. $Eu_2O_3$ (99.99%) and $TiO_2$ (99.99%) were used as the starting materials. The powders were ball milled at 350 rpm for 4 hours, and then calcinated at 1300 °C for 4h in air. The calcinated powder was then re-milled at 400 rpm for 4 hours to break any agglomerates and reduce the particle size. The powder was pressed as the green body under the pressure of 10 MPa for 15 min. The as-pressed bulk samples were treated at 1300 °C under different pressures of 5, 8, and 11 GPa for 30 min, respectively, using a high pressure machine with multi-anvils. The heating power is generated by the current pass through a graphite tube in a pyrophyllite block, which holds the sample during the experiments, and the heating rate is about 20 °C/min. The annealing experiments was using a box furnace, and the heating rate is 3 °C/min for all the samples.

X-ray diffraction (XRD) patterns for the samples were obtained with an X-ray diffractometer (X'pert Pro, PANalytical, Netherlands) using CuKa radiation. The microstructures of samples were observed using a scanning electron microscope (SEM, FEI Quanta 250, USA). The samples for SEM were polished, and then thermally etched at a temperature of 800 °C for 15 min to reveal their grain structures. The lattice structure was revealed by a transmission electron microscopy (TEM, FEI Tecnai $G^2$ F20 S-TWIN). Surface morphology of the ceramic was determined by atomic force microscopy (AFM) in contact mode, and the ferroelectric domain pattern was studied by a piezoresponse force microscopy (PFM) using a modified commercial atomic force microscope (Multimode, Nanoscope IIIa, Digital Instruments). Local piezoresponse loops were measured using a modified spectroscopic tool of PFM,[35, 45] by plotting the phase ɸ as a function of a DC voltage from -200 V to +200 V. Fired-on silver paste was used to make Electrodes for electrical property measurements. The frequency dependence of the dielectric constants and losses were measured using a Precision Impedance Analyzer (Agilent, 4294A, Hyogo, Japan). Samples for piezoelectric measurements were poled under various DC electric fields (20 - 30 kV/mm) in silicone oil at a temperature of 120 °C.[1, 27] We increased the electrical field gradually from 10 kV/mm until the sample was electrical breakdown to obtain high piezoelectric constant, $d_{33}$. The $d_{33}$ was measured using a quasi-static $d_{33}$ meter (CAS, ZJ-3B, Institute of Acoustics Chinese Academy of Sciences Beijing, China), with the instrument precision of 0.1 pC/N.[27, 30] To confirm the small $d_{33}$ is not an experiment error, both sides of the sample was measured and the $d_{33}$ are positive and negative on each side with the same absolute values. The $d_{33}$ is zero on the un-poled samples as measured. The thermal depoling behavior was investigated by holding the samples at a fixed temperature for 4 hours, then measuring their piezoelectric constant after cooling.

The calculations were carried out with density functional theory (DFT) in combination with the Vanderbilt-type ultrasoft pseudo-potential and a plane-wave expansion of the wave functions incorporated in the CASTEP code. The self-consistent ground state of the system was determined by using a band-by-band conjugate gradient technique to minimize the total energy of the system with respect to the plane-wave coefficients. The electronic wave functions were obtained by the Pulay density-mixing scheme and the structures were optimized by the BFGS method. The generalized gradient approximation (GGA) proposed by Perdew, Burke, and Ernzerhof (PBE)[46] was employed. The cutoff energy of plane waves was set to 480.0 eV and Brillouin zone sampling was performed by using the Monkhost-Pack[47] scheme with a k-point grid of 6 × 6 × 6, which were determined to ensure the convergence of total energies.

## Acknowledgements

This work was supported by the LSD project (Grant No.2016Z-04), Dean Fund of CAEP (Grant No. YZJJLX2016001), CSS project (Grant No. YK2015-0602006).

## Author Contributions

Z. G. conceived the idea and guided all the project. Z. G., Y. L. and C. L. completed the main manuscript text. Y. M. prepared figures 2-3. Y. X. and L. F. prepared figure 1. Q. H., G. L. and J. Y. prepared figure 4. H. H. and D. H. supported the high pressure experiments. L. Y. contributed the simulation works. All authors reviewed the manuscript.

## Additional Information

Competing financial interests: The authors declare no competing financial interests.

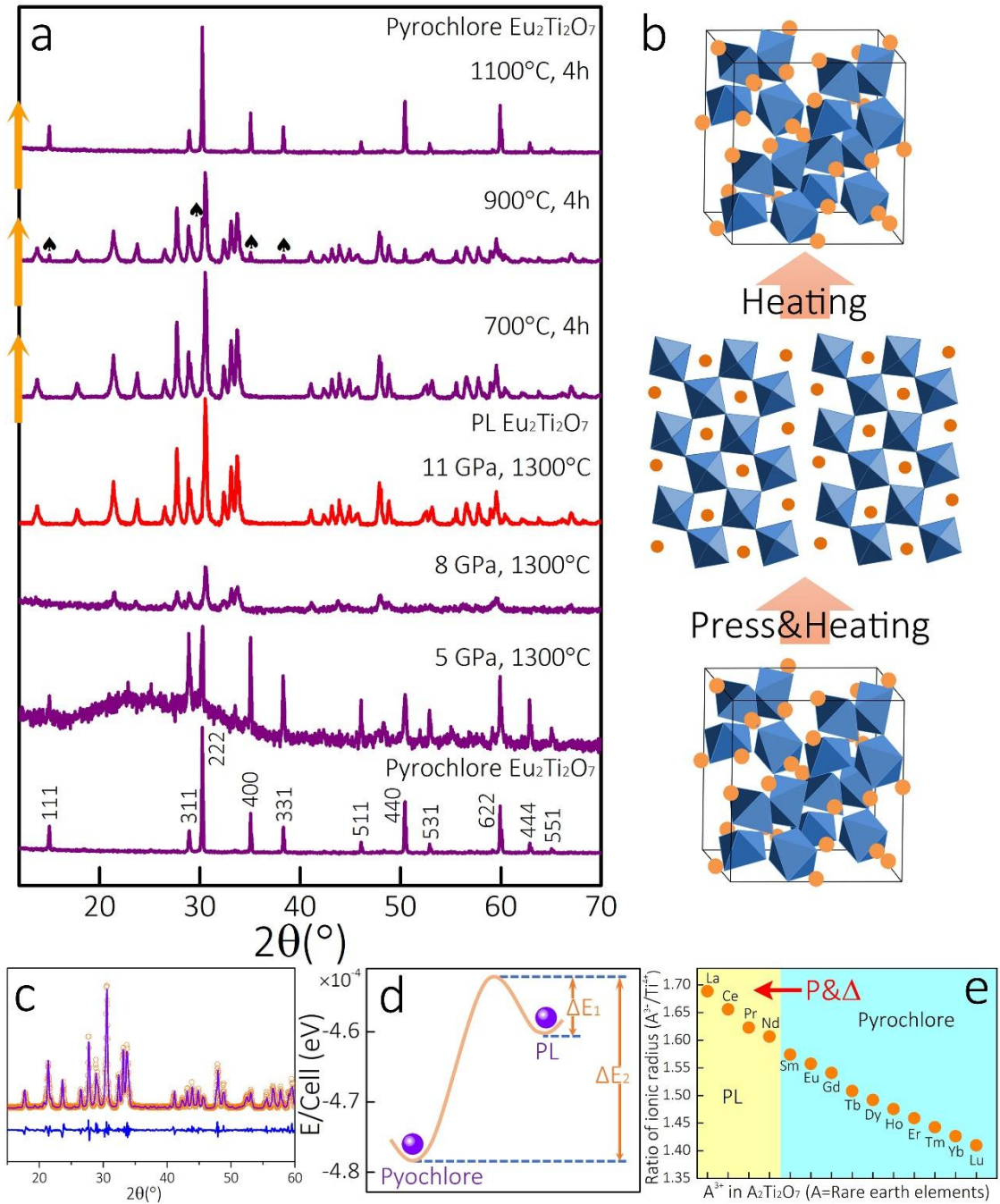

Figure 1. (a) The XRD patterns of the synthesized pyrochlore $Eu_2Ti_2O_7$ sample; pyrochlore samples treated at 1300 °C under different pressures of 5, 8, and 11 GPa, respectively; and the PL $Eu_2Ti_2O_7$ annealed at high temperatures of 700, 900 and 1100 °C continuously. The order of the annealing treatments is shown as the orange arrows on the left axis. (b) Overview of the structure change of $Eu_2Ti_2O_7$ during these processes, in which the blue polyhedrons represent $TiO_6$ octahedron, and the orange spheres represent $Eu^{3+}$ cations. With the treatment of 11 GPa and 1300 °C, the pyrochlore $Eu_2Ti_2O_7$ transfer into PL structure, and the PL structure could be stable above 700 °C. After 900 °C annealing, the XRD pattern show the presence of some pyrochlore $Eu_2Ti_2O_7$ highlighted by the spades (♠), and the structure change back to pyrochlore at 1100 °C completely. (c) Rietveld refinement for PL $Eu_2Ti_2O_7$. The difference curve (bottom, blue) indicates the level of agreement between the calculated (purple) and observed patterns ($R_{wp}$ = 18.9). (d) The cell energy calculated

by first principles simulation for pyrochlore $Eu_2Ti_2O_7$ and PL $Eu_2Ti_2O_7$ at 0 GPa, in which the energy of PL $Eu_2Ti_2O_7$ was calculated based on the $La_2Ti_2O_7$ structure, due to the atomic site occupation is not available for PL $Eu_2Ti_2O_7$. (e) The ratio of ionic radius of the $A_2Ti_2O_7$ compounds (A=Rare earth elements). When the $A^{3+}/Ti^{4+}$ ionic radius ratio is small, the structure is pyrochlore and when $A^{3+}/Ti^{4+}$ ionic radius ratio is large, the structure is PL. The pressure could induce the pyrochlore change into PL structure.

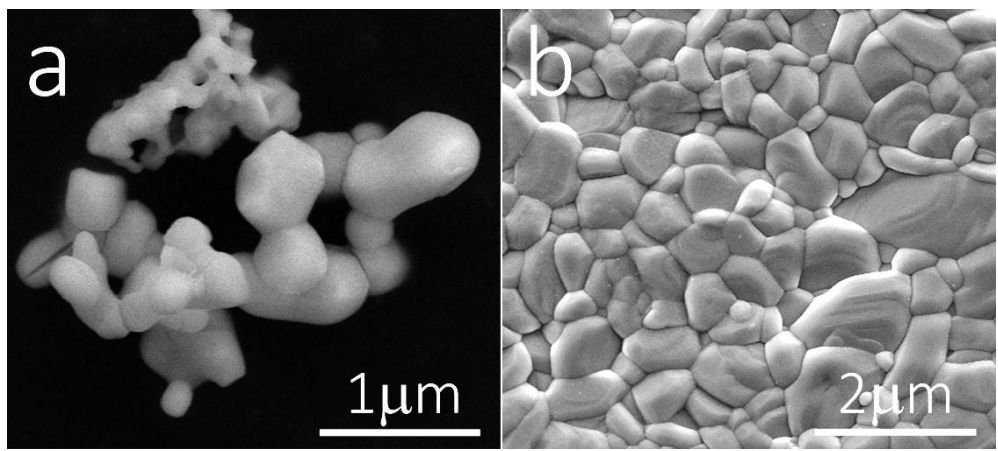

Figure 2. Scanning electron microscope micrograph of (a) the pyrochlore $Eu_2Ti_2O_7$ powder after ball milling and (b) the ceramic sintered at 11 GPa, 1300 °C, which is PL structure based on XRD.

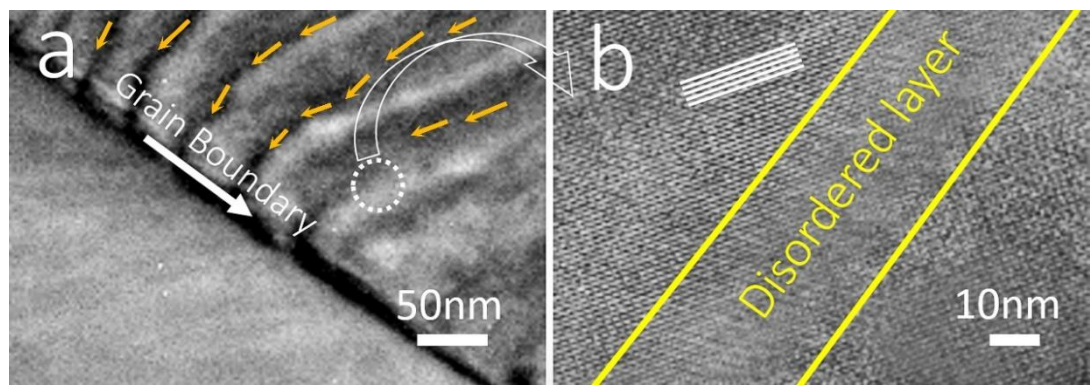

Figure 3. (a–b) HR-TEM images of $Eu_2Ti_2O_7$ ceramics with PL structure in two scales. The (b) image is the area in (a) marked by the circle with dashed line.

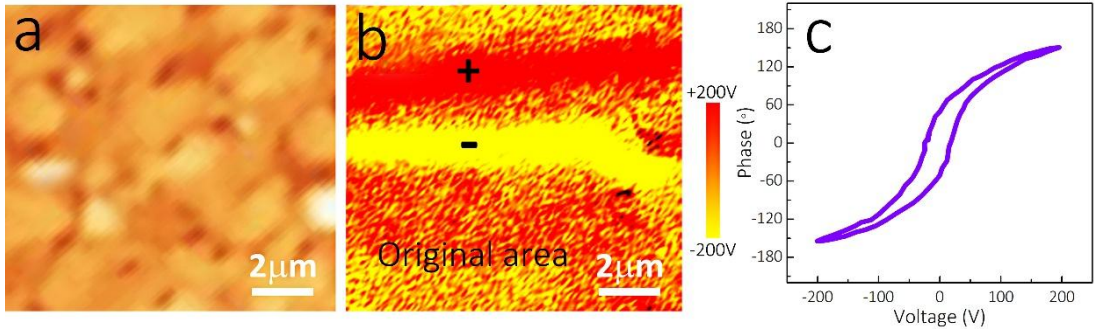

Figure 4. (a) AFM image of PL structure $Eu_2Ti_2O_7$ ceramic; and simultaneously recorded (b) characteristic PFM image, when a belt area has been polarized (red region) and another belt area has been reverse polarized (yellow region) by applying +200 V and -200 V, respectively. (c) Phase piezoresponse loop of the thin PL structure $Eu_2Ti_2O_7$ ceramic.

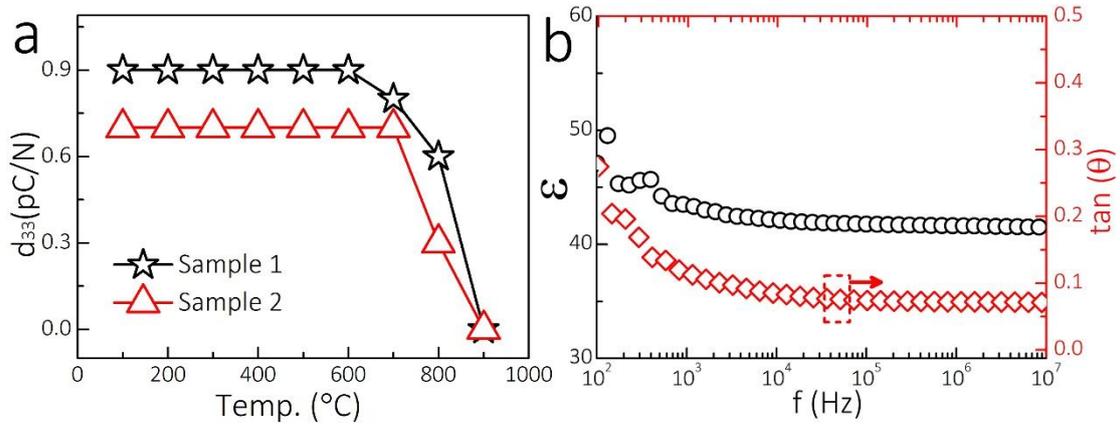

Figure 5. (a) Effect of thermal annealing on piezoelectric properties ($d_{33}$) of the poled PL structure $Eu_2Ti_2O_7$ ceramic. (b) Frequency dependence of dielectric constant and loss (tan(θ)) of PL structure $Eu_2Ti_2O_7$ ceramic.